# Generation and conversion of vortex light in coherently prepared hydrogen atoms and hydrogen-like ions

Fan Meng, Guo-Feng Zhang*
*School of Physics, Beihang University, Beijing 102206, China*

We propose a theoretical scheme for generating and converting optical vortices in a coherently prepared three-level Λ-type atomic medium. An azimuthally varying relative phase between the two lower states imparts additional phase winding to the generated field, yielding $l_2 = l_1 + m$ in the ideal model, where $l_1$ is the input topological charge and $m$ the coherence winding number. Thus, a Gaussian input can generate a vortex, while an incident vortex can be converted to a different orbital angular momentum (OAM) mode. Analytical results and numerical examples using hydrogen transition parameters illustrate both processes. Under resonant conditions, the model predicts strongly suppressed absorption during subsequent propagation as the bright-state component is attenuated and electromagnetically induced transparency (EIT) emerges. Spatial coherence is assumed as an initial condition; its detailed experimental preparation remains open. This scheme offers a potential framework for controlling vortex light through spatial atomic coherence.

## 1. Introduction

In recent years, vortex light has attracted considerable attention owing to its distinctive spatial phase structure and unique physical properties. Vortex light is characterized by a helical wavefront with an azimuthally varying phase factor $e^{il\varphi}$, where $l$ denotes the topological charge and $\varphi$ is the azimuthal angle in the transverse plane. Such a phase structure gives rise to a phase singularity, typically accompanied by a central intensity minimum, and exhibits topological robustness under continuous perturbations that preserve the singularity structure. For a pure orbital angular momentum (OAM) eigenmode with an integer topological charge $l$, each photon carries OAM of $l\hbar$ [1]. Orthogonal OAM modes offer an additional degree of freedom for high-dimensional optical information encoding and manipulation. These distinctive features have stimulated extensive research into vortex light and enabled a broad range of applications, particularly in quantum information storage, transmission, and processing [2-7], as well as optical tweezers [8], optical metrology [9], laser processing [10], and microscopic imaging [11].

Vortex beams are commonly generated using optical elements placed along the light propagation path [12]. Representative techniques include geometric-optics-based mode converters [13], spiral phase plates [14], spatial light modulators [15], and q-plates [16]. Although these techniques are well established, they may require specialized optical components and precise fabrication or alignment, increasing experimental complexity and cost. In addition, nonlinear optical processes provide an alternative route for conversion between distinct OAM modes [17,18], although their efficiencies can be limited by phase matching conditions and pump power. More recently, significant advances in vortex generation and control have been achieved using engineered photonic lattices. Li et al. generated multiple vortex-antivortex pairs in momentum space by propagating a focused Gaussian probe beam through honeycomb and hexagonal photonic lattices, with the resulting vortex configurations determined by the symmetry of the evolving wave packet [19]. Zhang et al.

* Corresponding author. E-mail:gf1978zhang@buaa.edu.cn

demonstrated spin-to-OAM conversion in non-Hermitian staggered photonic graphene, where the sign of the output OAM could be reversed by varying either the input circular polarization or the lattice staggering [20]. These studies establish important experimental routes to vortex generation and control.

Coherent light-matter interactions in atomic media provide a versatile framework for the generation and manipulation of structured and quantum light, with coherent optical dressing offering an effective means of controlling atomic resonances and nonlinear optical responses. Experimental advances have demonstrated the potential of such interactions for vortex light generation, OAM transfer, and the generation of nonclassical states of light. In particular, Zhang et al. experimentally demonstrated vortex six-wave mixing in a rubidium atomic ensemble, where the OAM carried by an incident probe beam was transferred to the generated optical field, accompanied by dressing controlled spatial modulation [21]. Li et al. subsequently demonstrated the generation of continuous-mode time–energy entangled W-class triphotons through spontaneous six-wave mixing (SSWM), in which multiple dressed resonance channels contribute to the generation process and the interference between different emission pathways gives rise to temporal quantum interference [22]. These studies highlight the roles of atomic coherence and optical dressing in tailoring nonlinear optical responses and controlling the spatial and quantum properties of generated light. They also underscore the potential of coherently controlled atomic media for structured and nonclassical light generation, motivating further exploration of alternative mechanisms for vortex light generation and OAM conversion. In this context, the scheme proposed below offers a related yet complementary approach based on spatially dependent atomic coherence.

In this paper, we develop a theoretical scheme for the direct generation of vortex light from a Gaussian input and conversion between distinct OAM modes in an effective three-level $\Lambda$ medium. Specifically, the two lower states are assumed to possess a spatially dependent coherence before the optical conversion process. The azimuthal winding of this coherence contributes an additional phase to the generated field, yielding the topological charge relation $l_2 = l_1 + m$ between the input field $\Omega_1$ and the generated field $\Omega_2$, where $m$ characterizes the prescribed coherence winding. Gaussian-input vortex generation and optical control of vortex chirality have already been demonstrated experimentally in three-level EIT atomic systems [19,20]. Our model considers a different phase-transfer pathway: the prescribed azimuthal winding of the lower-state coherence couples the input field to the generated field, without invoking a periodic photonic lattice or photonic spin-orbit coupling as the vortex generation mechanism. For $m = 0$, the generated field retains the input topological charge, recovering the single-input, spatially uniform coherence case considered in Ref. [23]; nonzero $m$ instead permits Gaussian to vortex generation and conversion between different OAM modes. We illustrate this mechanism numerically using hydrogen transition parameters and discuss the rotation properties of hydrogen atoms and hydrogen-like ions as a formal description of relative phase control. Within the ideal model, absorption occurs predominantly near the medium entrance and is strongly suppressed during subsequent propagation under the assumed EIT conditions. The present study is a theoretical proposal whose propagation predictions assume that the required spatial coherence has been established and maintained; its detailed experimental preparation remains unresolved.

## 2. Theoretical Model

We consider a three-level Λ-type atomic system, as illustrated in Fig. 1. The state $|e\rangle$ denotes the excited level, whereas $|g_1\rangle$ and $|g_2\rangle$ correspond to the two lower states. We investigate the propagation of two optical fields with Rabi frequencies $\Omega_1$ and $\Omega_2$ in a Λ-type medium, where the coupling field $\Omega_1$ drives the $|g_1\rangle \leftrightarrow |e\rangle$ transition, whereas the probe field $\Omega_2$ drives the $|g_2\rangle \leftrightarrow |e\rangle$ transition. In the appropriate rotating frame, the Hamiltonian of the system can be written as

$$H = -\hbar\Delta_1 |g_1\rangle\langle g_1| - \hbar\Delta_2 |g_2\rangle\langle g_2| - \hbar\left(\Omega_1 |e\rangle\langle g_1| + \Omega_2 |e\rangle\langle g_2| + \text{H.c.}\right), \tag{1}$$

where $\Omega_1 = \mu_{eg_1} E_1 / 2\hbar$ and $\Omega_2 = \mu_{eg_2} E_2 / 2\hbar$ denote the Rabi frequencies, $\mu_{eg_1}$ and $\mu_{eg_2}$ are the electric dipole matrix elements, and $E_1$ and $E_2$ are the amplitudes of the corresponding optical fields. The detunings are defined as $\Delta_1 = \omega_{eg_1} - \omega_1$ and $\Delta_2 = \omega_{eg_2} - \omega_2$, representing the frequency offsets between the atomic transitions and the optical fields.

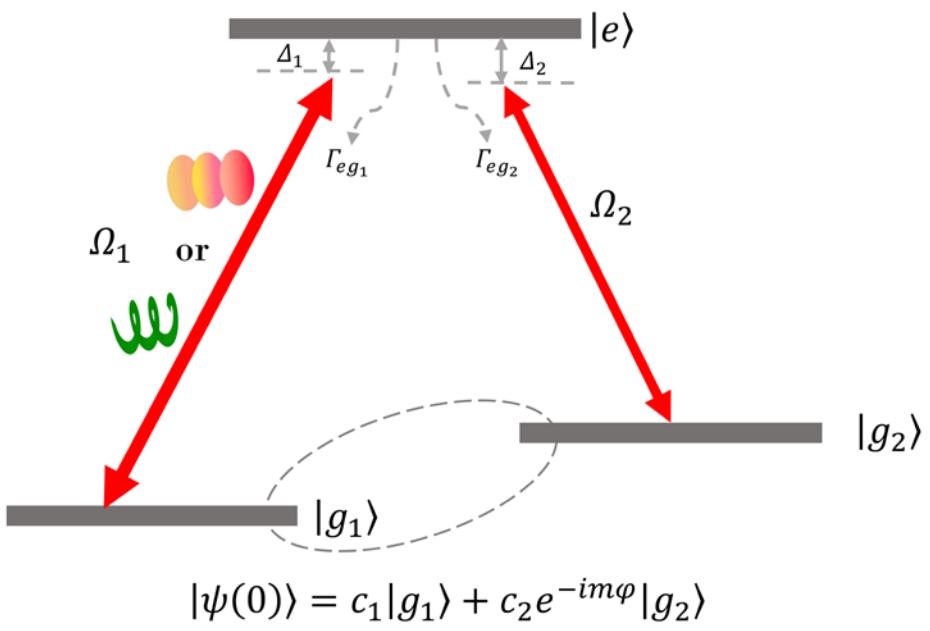

**Fig. 1.** Effective three-level Λ configuration for coherence-mediated optical field conversion. The excited state is denoted by $|e\rangle$, while $|g_1\rangle$ and $|g_2\rangle$ denote the two lower state levels. The coupling field $\Omega_1$ drives the transition between $|g_1\rangle$ and $|e\rangle$, whereas the probe $\Omega_2$ drives the transition between $|g_2\rangle$ and $|e\rangle$. The detunings of the two transitions are denoted by $\Delta_1$ and $\Delta_2$, while the decay rates from the excited state to the corresponding lower states are denoted by $\Gamma_{eg_1}$ and $\Gamma_{eg_2}$. The two lower states of the medium are prepared in a spatially dependent coherent superposition, $|\psi(0)\rangle = c_1|g_1\rangle + c_2 e^{-im\varphi}|g_2\rangle$.

The dynamical evolution of the system is governed by the Liouville equation, which can be written as

$$\frac{\partial}{\partial t}\rho = L_\rho - \frac{i}{\hbar}[H, \rho], \tag{2}$$

where the first term on the right-hand side accounts for the damping and relaxation processes, whereas the second term describes the coherent dynamical evolution of the system.

By substituting Eq. (1) into Eq. (2), we obtain the explicit equations of motion for this three-level system as follows:

$$\frac{\partial}{\partial t}\rho_{eg_1} = -\left(\frac{\Gamma_{eg_1} + \Gamma_{eg_2}}{2} + i\Delta_1\right)\rho_{eg_1} + i\Omega_1\left(\rho_{g_1g_1} - \rho_{ee}\right) + i\Omega_2\rho_{g_2g_1}, \tag{3}$$

$$\frac{\partial}{\partial t}\rho_{eg_2} = -\left(\frac{\Gamma_{eg_1}+\Gamma_{eg_2}}{2}+i\Delta_2\right)\rho_{eg_2} + i\Omega_2\left(\rho_{g_2g_2}-\rho_{ee}\right)+i\Omega_1\rho_{g_1g_2}, \tag{4}$$

where $\Gamma_{eg_1}$ and $\Gamma_{eg_2}$ denote the decay rates from the upper level to the corresponding lower levels, respectively. For simplicity, we assume $\Gamma_{eg_1}=\Gamma_{eg_2}=\Gamma_e/2$.

The propagation dynamics of the light fields $\Omega_1$ and $\Omega_2$ within the medium are described by the Maxwell-Bloch equations, expressed as follows:

$$\left(\frac{1}{c}\cdot\frac{\partial}{\partial t}+\frac{\partial}{\partial z}\right)\Omega_1 = \frac{i}{2k_1}\nabla_\perp^2\Omega_1 + i\frac{\Gamma_{eg_1}\alpha_1}{2L}\rho_{eg_1}, \tag{5}$$

$$\left(\frac{1}{c}\cdot\frac{\partial}{\partial t}+\frac{\partial}{\partial z}\right)\Omega_2 = \frac{i}{2k_2}\nabla_\perp^2\Omega_2 + i\frac{\Gamma_{eg_2}\alpha_2}{2L}\rho_{eg_2}, \tag{6}$$

where $\alpha_1$ and $\alpha_2$ denote the optical densities associated with the two optical fields, $k_1=\omega_1/c$ and $k_2=\omega_2/c$ are the corresponding wave vectors, and $L$ is the medium length. For simplicity, we set $\alpha_1=\alpha_2=\alpha$.

We construct a spatially dependent coherent superposition state, assuming that the atoms are initially prepared in the state

$$\left|\psi(0)\right\rangle = c_1\left|g_1\right\rangle + c_2 e^{-im\varphi}\left|g_2\right\rangle, \tag{7}$$

where $\varphi$ denotes the azimuthal angle around the $z$-axis, $m$ represents the winding quantum number. and $c_1$ and $c_2$ are the normalization constants. For $m\neq 0$, the coherence is spatially varying in phase. For $m=0$, the coherence is spatially uniform. This spatial coherence is prescribed as an initial condition for the propagation calculation; its detailed experimental preparation remains unresolved.

Under the weak-interaction approximation between the light field and the atomic ensemble, we may set $\rho_{ee}\approx 0$, $\rho_{g_1g_1}\approx|c_1|^2$, $\rho_{g_2g_2}\approx|c_2|^2$, and $\rho_{g_1g_2}\approx c_1c_2^*e^{im\varphi}$ to a good approximation. However, when the medium is sufficiently long, the propagation of the light field evolves toward a steady state, in which case the weak-interaction assumption can be relaxed [24]. We consider the steady state regime, in which the system remains time-independent. By substituting the solutions of Eqs. (3) and (4) into the Maxwell-Bloch equations, Eqs. (5) and (6), we obtain a coupled set of equations governing the propagation of the light fields in the medium:

$$\frac{\partial\Omega_1}{\partial z} = \frac{i}{2k_1}\nabla_\perp^2\Omega_1 - iX_1(|c_1|^2\Omega_1 + c_1^*c_2e^{-im\varphi}\Omega_2), \tag{8}$$

$$\frac{\partial\Omega_2}{\partial z} = \frac{i}{2k_2}\nabla_\perp^2\Omega_2 - iX_2(c_1c_2^*e^{im\varphi}\Omega_1 + |c_2|^2\Omega_2), \tag{9}$$

where $X_1=\Gamma_{eg_1}\alpha_1/\left[2L\left(i\Gamma_{eg_1}-\Delta_1\right)\right]$, $X_2=\Gamma_{eg_2}\alpha_2/\left[2L\left(i\Gamma_{eg_2}-\Delta_2\right)\right]$.

We impose the boundary conditions $\Omega_1(r,\varphi,0)=f_{in}(r)e^{il\varphi}$ and $\Omega_2(r,\varphi,0)=0$ at the entrance of the medium ($z=0$). By solving Eqs. (8) and (9), the corresponding steady state solutions for the propagation of the light fields in the medium are obtained

as follows:

$$\Omega_1(r,\varphi,z)=\exp\left[z\cdot\hat{A}_{l,m}^{(r)}\right]_{11} f_{in}(r)e^{il\varphi}, \tag{10}$$

$$\Omega_2(r,\varphi,z)=\exp\left[z\cdot\hat{A}_{l,m}^{(r)}\right]_{21} f_{in}(r)e^{i(l+m)\varphi}, \tag{11}$$

where $\hat{A}_{l,m}^{(r)}=\begin{pmatrix}\frac{i}{2k_1}\hat{B}_l^{(r)}-iX_1|c_1|^2 & -iX_1c_1^*c_2 \\ -iX_2c_1c_2^* & \frac{i}{2k_2}\hat{B}_{l+m}^{(r)}-iX_2|c_2|^2\end{pmatrix}$ denotes the radial propagation operator, $\hat{B}_n^{(r)}=\frac{\partial^2}{\partial r^2}+\frac{1}{r}\frac{\partial}{\partial r}-\frac{n^2}{r^2}$ represents the radial diffraction operator, and $f_{in}(r)=\Omega_0\left(\frac{r}{w_0}\right)^{|l|}e^{-r^2/w_0^2}$ characterizes the transverse profile of the initial input field. A detailed derivation is provided in the Appendix.

We consider the resonant condition $\Delta_1=\Delta_2=0$, corresponding to the case in which the light fields are exactly resonant with the atomic transitions. Under this condition, $X_1=1/(2iL_{abs})$, $X_2=1/(2iL_{abs})$, where $L_{abs}=L/\alpha$ denotes the absorption length of the medium. When the optical density of the medium is sufficiently large ($\alpha\gg 1$), the absorption length occupies only a small fraction of the medium, i.e., $L_{abs}\ll L$. In the region $z\gg L_{abs}$, the leading bright-state contribution, which decays as $\exp\left[-z/(2L_{abs})\right]$ along the propagation direction, is strongly suppressed. The optical fields subsequently propagate predominantly in the dark subspace with weak residual absorptive loss, while their transverse profiles continue to evolve because of diffraction. This behavior arises from the establishment of the EIT condition, under which the atoms are driven into a dark state [25]. The normalized field intensity is defined as

$$P_j(z)=\frac{\int_0^{2\pi}\int_0^{\infty}\left|\Omega_j(r,\varphi,z)\right|^2 rdrd\varphi}{\int_0^{2\pi}\int_0^{\infty}\left|\Omega_1(r,\varphi,0)\right|^2 rdrd\varphi}\quad (j=1,2), \tag{12}$$

which characterizes the evolution of the relative intensities of the input field $\Omega_1$ and the generated field $\Omega_2$ during propagation through the medium. Figure 2 illustrates the evolution of the normalized input-field intensity $P_1$ and the generated field intensity $P_2$ as functions of the dimensionless propagation length $z/L$ under the conditions $\Delta_1=\Delta_2=0$, $\alpha_1=\alpha_2=30$, and $c_1=c_2=1/\sqrt{2}$. The results indicate that, as the interaction depth between the light field and the medium increases, a portion of the energy carried by the initial input field $\Omega_1$ is progressively transferred to the generated field $\Omega_2$. Simultaneously, this process is accompanied by OAM transfer, with the optical field gradually evolving towards a stable propagation regime once EIT is established.

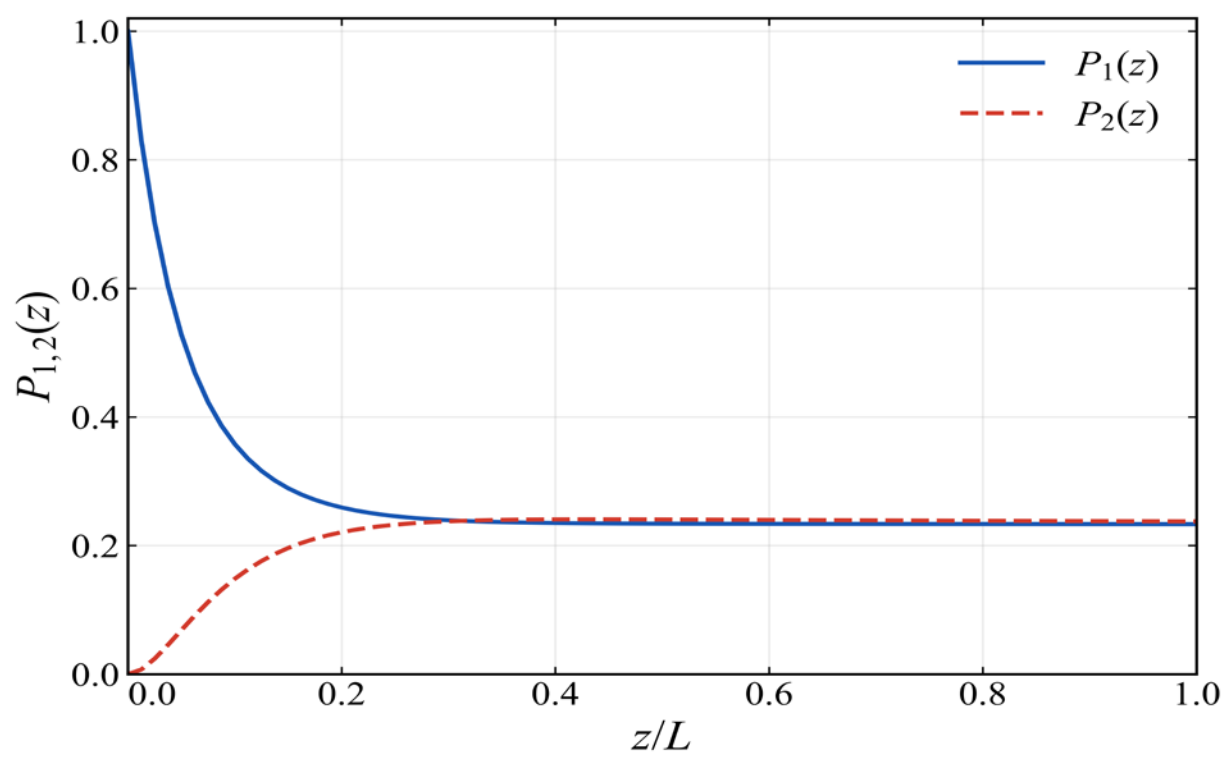

**Fig. 2.** Propagation of light fields in a Λ-type medium. The dependence of the relative intensities $P_1$ and $P_2$ on the dimensionless length $z/L$ when $\Delta_1=\Delta_2=0$, $\alpha_1=\alpha_2=30$, and $c_1=c_2=1/\sqrt{2}$.

Before discussing the analytical results, we first clarify the distinctions and connections between conventional Gaussian beams and vortex beams. In cylindrical coordinates, the general form of a Laguerre-Gaussian (LG) vortex beam can be written as:

$$\Omega_{LG}(r,\varphi,0)=f_{in}(r)e^{il\varphi}=\Omega_0\left(\frac{r}{w_0}\right)^{|l|}e^{-r^2/w_0^2}e^{il\varphi}, \tag{13}$$

where $\Omega_0$, $r$, $w_0$, and $l$ denote the intensity of the LG beam, the radial coordinate, the beam waist, and the topological charge index, respectively. Upon a single rotation around the phase singularity, the helical phase profile acquires a phase shift of $2\pi l$, which originates from the azimuthal phase factor $\exp(il\varphi)$ in the field expression. It has been theoretically established that such a vortex beam carries an OAM of $l\hbar$ per photon. In particular, when $l=0$, the vortex beam reduces to a Gaussian beam.

For a Gaussian beam, the field takes the form

$$\Omega_{Guass}(r,\varphi,0)=\Omega_0 e^{-r^2/w_0^2}, \tag{14}$$

First, we consider the case in which the initial incident field $\Omega_1$ is a Gaussian beam, i.e., $\Omega_1(r,\varphi,0)=\Omega_{Guass}(r,\varphi,0)$. From Eqs. (10) and (11), we find that the generated field $\Omega_2$ acquires a vortex phase structure, since $\Omega_2(r,\varphi,z)\sim\Omega_{Guass}(r,\varphi,0)e^{im\varphi}$. This indicates that the interaction of the Gaussian input field $\Omega_1$ with the medium directly generates the vortex field $\Omega_2$. This conversion is not realized through conventional optical elements inserted in the beam path, but rather through preparing the lower state of the medium in a tailored coherent superposition state $|\psi(0)\rangle=c_1|g_1\rangle+c_2e^{-im\varphi}|g_2\rangle$. One possible approach to preparing this superposition state is described in Sec. 3. This scheme thus provides a viable pathway for the direct generation of vortex light. Next, we consider the case where the initial incident field $\Omega_1$ is itself a vortex beam, i.e., $\Omega_1(0)=\Omega_{LG}(r,\varphi,0)=f_{in}(r)e^{il\varphi}$. According to Eqs. (10) and (11), the generated field $\Omega_2$ forms a new vortex mode carrying an OAM of $(l+m)\hbar$ per photon, since $\Omega_2(r,\varphi,z)\sim f_{in}(r)e^{i(l+m)\varphi}$. In this manner, a controlled conversion between distinct OAM states of the vortex field is realized. This process does not rely on multi-wave mixing schemes; instead, it is likewise enabled by preparing the lower state of the medium in the coherent superposition state $\left|\psi(0)\right\rangle = c_1\left|g_1\right\rangle + c_2 e^{-im\varphi}\left|g_2\right\rangle$. It is worth emphasizing that the azimuthal phases in Eqs. (10) and (11) lead to the topological charge relation $l_2 = l_1 + m$ between the input and generated fields. The additional winding $m$ originates from the prescribed lower-state coherence. For $m \neq 0$, a Gaussian input can generate a vortex, while a vortex input can be converted to a different charge. For $m = 0$, the generated field inherits the input winding within this model. This differs from the topological charge relation for the generation pathway employed in the vortex six-wave mixing experiment of Ref. [21], $l_s = l_1 + l_3 - l_{3'}$. For the Gaussian auxiliary fields used there, $l_3 = l_{3'} = 0$. The distinction lies in the spatial phase supplied by the prepared medium in the present model, rather than in the transfer of the vortex phase from the input fields through the six-wave mixing process.

In addition, it is useful to distinguish among confinement-induced energy discretization, coupling-induced spectral restructuring, and static or dynamically induced level splitting. These mechanisms are classified according to their physical origins and are not necessarily mutually exclusive. First, finite size confinement in a quantum dot imposes boundary conditions on the carrier wavefunctions, giving rise to size dependent discrete energy levels [26]. This mechanism is distinct from the confinement considered in the present model. Second, coupling mixes bare states and can produce avoided crossings. For two bare levels $E_a$ and $E_\beta$ coupled by an energy matrix element $g$, an ideal Hermitian model gives

$$E_+ = \left(E_a + E_\beta\right)/2 + \sqrt{\left(E_a - E_\beta\right)^2/4 + |g|^2}\,, \tag{15}$$

$$E_- = \left(E_a + E_\beta\right)/2 - \sqrt{\left(E_a - E_\beta\right)^2/4 + |g|^2}\,. \tag{16}$$

At bare resonance, their separation is $2|g|$. A nonzero coupling therefore produces an avoided crossing when the bare-level separation is varied through resonance. If symmetry forbids the coupling, a genuine crossing may remain. Avoided crossing describes the behavior of coupled eigenenergies as a parameter is varied; it is not synonymous with the existence of discrete bound levels. Periodic coupling in a lattice generally produces Bloch bands and gaps, with additional finite size discretization when appropriate [27]. Our propagation model does not contain an atomic lattice or intersite tunneling. Third, a static crystal field can split and mix atomic multiplets according to the local environment and its symmetry [28]. This is distinct from size-dependent confinement. Optical dressing instead results from coherent coupling by an oscillating electromagnetic field. The dressed eigenstates combine the coupled atomic states, and their rotating-frame energies depend on the drive strength and detuning [29]. Dressing is itself a form of level coupling and can exhibit avoided crossings; it is not independent of the preceding coupling mechanism. Static crystal-field splitting and dynamic optical dressing should nevertheless be distinguished by their physical origin. The present $\Lambda$ model comprises intrinsically discrete atomic levels coupled coherently by optical fields, but contains no quantum dot confinement potential, atomic lattice, or crystal-field interaction. At exact resonance, for fixed local field amplitudes and in the absence of relaxation, the $\Lambda$ system also admits a local dark-state description. Its local dark state can be written as

$$\left|D\right\rangle = \frac{\Omega_2\left|g_1\right\rangle - \Omega_1\left|g_2\right\rangle}{\sqrt{\left|\Omega_1\right|^2 + \left|\Omega_2\right|^2}}. \tag{17}$$

The corresponding local dressed state eigenenergies are $\varepsilon_+ = +\hbar\Omega_R$, $\varepsilon_D = 0$, and $\varepsilon_- = -\hbar\Omega_R$, where $\Omega_R = \sqrt{\left|\Omega_1\right|^2 + \left|\Omega_2\right|^2}$. The zero energy dark state has no excited state component. Its destructive interference accounts for the suppression of absorption, whereas the winding number of the spatial coherence determines the change in the vortex topological charge. Dark-state dressing and spatial winding therefore describe different aspects of the interaction. Absorption suppression after bright-component attenuation does not imply unit conversion efficiency, and diffraction can produce residual loss.

**3. Construction of the spatially dependent coherent superposition state**

We theoretically propose a scheme for the direct generation and conversion of vortex light in coherently prepared media. The medium is not arbitrarily prepared; rather, it requires that its lower state be initialized in a tailored coherent superposition state $\left|\psi(0)\right\rangle = c_1\left|g_1\right\rangle + c_2 e^{-im\varphi}\left|g_2\right\rangle$, where $m$ denotes the winding quantum number and $\varphi$ is the azimuthal angle around the $z$-axis. This coherent superposition state contains an intrinsic azimuthal phase factor $e^{-im\varphi}$, which is mapped onto the optical field through the atom-light interaction. As a consequence, the light field acquires the conjugate azimuthal phase factor $e^{im\varphi}$, thereby being modulated accordingly. By analyzing the wavefunctions of hydrogen atoms and hydrogen-like ions, we show that the desired spatially dependent coherent superposition can be formally constructed using appropriate rotation operators.

The stationary wave function of the hydrogen atom can be expressed as a product of a radial component and an angular component:

$$\phi_{nlm}(r,\theta,\varphi) = R_{nl}(r)Y_{lm}(\theta,\varphi), \tag{18}$$

where $R_{nl}(r) = N_{nl}\left[2r/(na_0)\right]^l \exp\left[-r/(na_0)\right] L_{n-l-1}^{2l+1}\left[2r/(na_0)\right]$ denotes the radial wave function, $N_{nl} = \sqrt{(2/na_0)^3\left\{(n-l-1)!/\left[2n(n+l)!\right]\right\}}$ is the normalization factor, and $a_0$ represents the Bohr radius. The term $L_{n-l-1}^{2l+1}\left[2r/(na_0)\right]$ corresponds to the associated Laguerre function, while $Y_{lm}(\theta,\varphi) = (-1)^m\sqrt{\left[(2l+1)(l-m)!\right]/\left[4\pi(l+m)!\right]}P_l^m(\cos\theta)\exp(im\varphi)$ denotes the spherical harmonic function, and $P_l^m(\cos\theta)$ is the associated Legendre polynomial.

Hydrogen-like ions are atomic species characterized by the presence of a single electron orbiting the nucleus. In the context of the wave function of hydrogen-like ions, owing to the difference in the electron reduced mass compared with hydrogen atoms, the radial wave function $R_{nl}(r)$ can be obtained by substituting the Bohr radius $a_0$ with $a/Z$, where $a$ denotes the effective Bohr radius of the hydrogen-like ion and $Z$ represents its nuclear charge. The angular part, described by spherical harmonics, remains unchanged. The angular part of the wavefunctions of hydrogen atoms and hydrogen-like ions contains an azimuthal phase factor $e^{im\varphi}$. However, the presence of this phase factor does not directly imply the formation of the desired spatially dependent coherent superposition, $\left|\psi(0)\right\rangle = c_1\left|g_1\right\rangle + c_2 e^{-im\varphi}\left|g_2\right\rangle$, since the azimuthal phase is an intrinsic part of the orbital wavefunction rather than an independently prescribed relative phase between the two lower states. Within the orbital approximation, the action of an appropriate rotation operator provides a formal means of introducing a relative phase between the relevant atomic states.

The rotation operator corresponding to a rotation about the $z$-axis is given by:

$$\hat{\Re} = \exp\left(-\frac{i}{\hbar}\varphi \hat{L}_z\right), \tag{19}$$

where $\varphi$ denotes the infinitesimal rotation angle about the $z$-axis, and $\hat{L}_z$ represents the $z$-component of the OAM operator.

By applying the rotation operator $\hat{\Re}$ to the wave function of a hydrogen atom and hydrogen-like ion, one obtains:

$$\hat{\Re}\phi_{nlm}(r,\theta,\varphi) = R_{nl}(r)\hat{\Re}Y_{lm}(\theta,\varphi) = e^{-im\varphi}\phi_{nlm}(r,\theta,\varphi). \tag{20}$$

Based on Eq. (20), the coherent superposition state given in Eq. (7) can be constructed. For instance:

$$\left|\psi\right\rangle = c_1\left|\phi_{210}\right\rangle + c_2\hat{\Re}\left|\phi_{211}\right\rangle = c_1\left|\phi_{210}\right\rangle + c_2 e^{-i\varphi}\left|\phi_{211}\right\rangle, \tag{21}$$

$$\left|\psi\right\rangle = c_1\left|\phi_{210}\right\rangle + c_2\hat{\Re}\left|\phi_{21-1}\right\rangle = c_1\left|\phi_{210}\right\rangle + c_2 e^{i\varphi}\left|\phi_{21-1}\right\rangle, \tag{22}$$

$$\left|\psi\right\rangle = c_1\left|\phi_{320}\right\rangle + c_2\hat{\Re}\left|\phi_{322}\right\rangle = c_1\left|\phi_{320}\right\rangle + c_2 e^{-i2\varphi}\left|\phi_{322}\right\rangle, \tag{23}$$

$$\cdots$$

Although the desired coherent superposition cannot be directly obtained from the unmodified wavefunctions of hydrogen atoms and hydrogen-like ions, it can be formally constructed by applying appropriate rotation operators to these wavefunctions. Within the orbital approximation, these rotation relations provide a formal description of relative phase control in hydrogen atoms and hydrogen-like ions. However, these relations should not be interpreted as a complete experimental preparation protocol. The physical implementation of the required spatial control and the preservation of the resulting coherence therefore remain open challenges. Moreover, the desired coherent superposition may also be achievable in other media, a possibility that merits further investigation.

## 4. Numerical results

Numerical calculations based on an effective three-level $\Lambda$ model of hydrogen illustrate the possibility of directly generating and converting vortex light, provided that the two lower states are prepared in a spatially dependent coherent superposition. Within the hydrogen atomic, we choose the excited state $\left|e\right\rangle$ to correspond to the $\left|3S_{1/2}\right\rangle$ state, while the two lower states $\left|g_1\right\rangle$ and $\left|g_2\right\rangle$ are assigned to the $\left|2P_{1/2}\right\rangle$ and $\left|2P_{3/2}\right\rangle$ states, respectively. The transition wavelength for $\left|3S_{1/2}\right\rangle \leftrightarrow \left|2P_{1/2}\right\rangle$ is 656.275181nm, whereas that for $\left|3S_{1/2}\right\rangle \leftrightarrow \left|2P_{3/2}\right\rangle$ is 656.290944nm. Therefore, the corresponding superposition states $\left|\psi\right\rangle = c_1\left|\phi_{210}\right\rangle + c_2 e^{-i\varphi}\left|\phi_{211}\right\rangle$ (with $\left|e\right\rangle \rightarrow \left|3S_{1/2}, m=0\right\rangle$ , $\left|g_1\right\rangle \rightarrow \left|2P_{1/2}, m=0\right\rangle$ , and $\left|g_2\right\rangle \rightarrow \left|2P_{3/2}, m=1\right\rangle$ ) and $\left|\psi\right\rangle = c_1\left|\phi_{210}\right\rangle + c_2 e^{i\varphi}\left|\phi_{21-1}\right\rangle$ (with $\left|e\right\rangle \rightarrow \left|3S_{1/2}, m=0\right\rangle$ , $\left|g_1\right\rangle \rightarrow \left|2P_{1/2}, m=0\right\rangle$ , and $\left|g_2\right\rangle \rightarrow \left|2P_{3/2}, m=-1\right\rangle$) can be constructed via the action of a rotation operator.

We first consider the scenario in which a Gaussian beam generates a vortex beam. After the two lower states of the three-level Λ-type hydrogen system are prepared in the coherent superposition state described above, a Gaussian input beam can directly generate a vortex beam. We define the transverse profile of the Gaussian field $\Omega_1$ at the medium entrance ($z=0$) as $\Omega_1(r,\varphi,0) = \Omega_{Guass}(r,\varphi,0) = \Omega_0 e^{-r^2/w_0^2}$ , where $\Omega_0$ denotes the Gaussian beam amplitude, $r$ is the radial coordinate, and $w_0$ is the incident beam waist ($w_0 = 1\text{mm}$). When the incident Gaussian field $\Omega_1$ interacts with the coherently prepared medium, a vortex field $\Omega_2$ is generated directly. The evolution of the two optical fields, $\Omega_1$ and $\Omega_2$, inside the medium is governed by Eqs. (10) and (11). Figure 3a presents a schematic illustration of vortex beam generation from a Gaussian beam. Figure 3b shows the corresponding numerical results. The first row gives the intensity and phase distributions of the incident Gaussian field $\Omega_1$ at $z=0$; the second row illustrates the medium prepared in the target coherent superposition state; and the third row displays the intensity and phase distributions of the generated vortex field $\Omega_2$ at $z=L$. Specifically, when the incident field $\Omega_1$ is a Gaussian beam with zero topological charge ($l_1 = 0$), its interaction with a medium prepared in the spatially dependent coherent superposition $\left|\psi\right\rangle = c_1\left|\phi_{210}\right\rangle + c_2 e^{-i\varphi}\left|\phi_{211}\right\rangle$ generates a vortex field $\Omega_2$ carrying a topological charge of $l_2 = 1$. By contrast, when the medium is prepared in the coherent superposition $\left|\psi\right\rangle = c_1\left|\phi_{210}\right\rangle + c_2 e^{i\varphi}\left|\phi_{21-1}\right\rangle$ the generated vortex field carries a topological charge of $l_2 = -1$. These results demonstrate that the topological charge of the generated vortex field can be controlled by tailoring the spatial phase structure of the prepared atomic coherence.

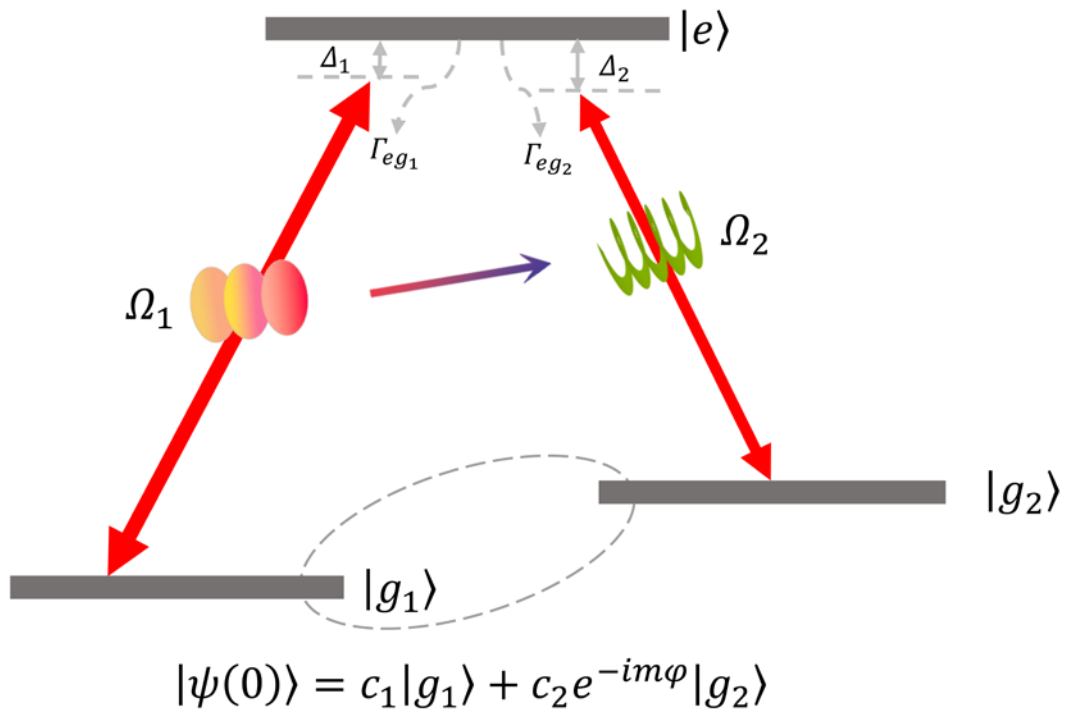

(a)

<table>
<tr><td>$\Omega_1$<br>Input</td><td colspan="2">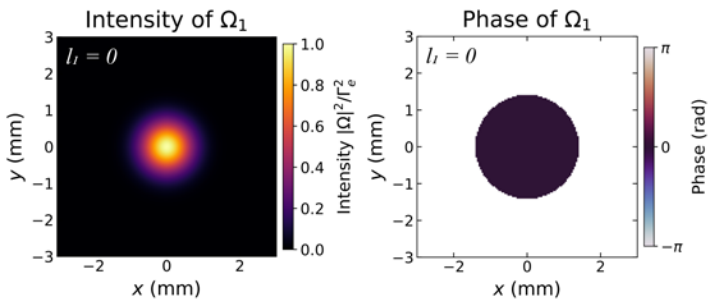</td></tr>
<tr><td>Superposition state</td><td>$\left|\psi\right\rangle = c_1\left|\phi_{210}\right\rangle + c_2 e^{-i\varphi}\left|\phi_{211}\right\rangle$</td><td>$\left|\psi\right\rangle = c_1\left|\phi_{210}\right\rangle + c_2 e^{i\varphi}\left|\phi_{21-1}\right\rangle$</td></tr>
<tr><td>$\Omega_2$<br>Output</td><td>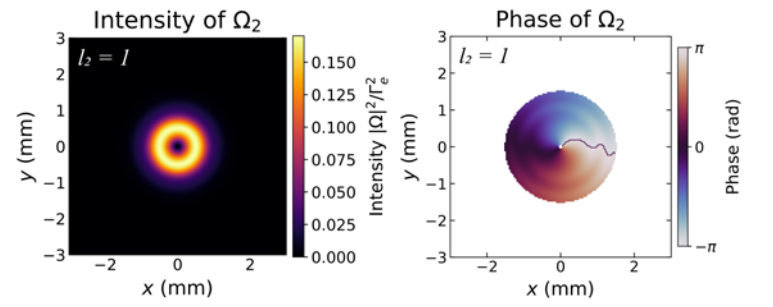</td><td>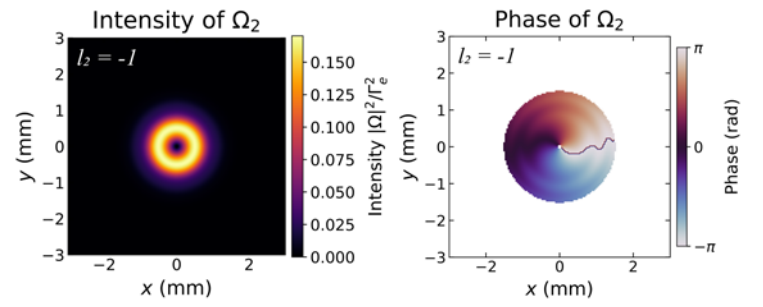</td></tr>
</table>

(b)

**Fig. 3.** Direct generation of a vortex beam from a Gaussian beam. (a) Schematic illustration of a vortex beam induced by a Gaussian beam in a three-level Λ-type hydrogen atomic system. (b) Numerical results. The first row shows the intensity and phase distributions of the Gaussian beam, whose initial transverse profile is given by $\Omega_1(r,\varphi,0)=\Omega_0 e^{-r^2/w_0^2}$ with $w_0 = 1\text{mm}$ .The second row depicts the preparation of the atomic lower state into the corresponding coherent superposition state of the medium. The third row presents the intensity and phase distributions of the generated vortex field after propagation over a distance $z = L$ . All other parameters are the same as those used in Fig. 2.

Next, we consider the conversion between different OAM vortex modes. Similarly, after preparing the two lower states in the three-level hydrogen system in the corresponding coherent superposition, conversion between different OAM vortex beams can be achieved. We assume that at the medium entrance ($z = 0$), the incident field $\Omega_1$ is a vortex beam with transverse profile $\Omega_1(r,\varphi,0)=\Omega_{LG}(r,\varphi,0)=\Omega_0(r/w_0)^{|l|} e^{-r^2/w_0^2} e^{il\varphi}$, where all parameters are defined and assigned values as described above. Upon interacting with the coherently prepared medium, the incident vortex field $\Omega_1$ generates a vortex field $\Omega_2$ carrying a different OAM mode. The evolution of the two optical fields, $\Omega_1$ and $\Omega_2$, within the medium is governed by Eqs. (10) and (11). Figure 4a presents a schematic illustration of the conversion between different OAM vortex beams. Figure 4b presents the corresponding numerical results. The first row depicts the intensity and phase distributions of the input vortex field $\Omega_1$, the second row illustrates the medium prepared in the target coherent superposition state, and the third row shows the intensity and phase distributions of the generated vortex field $\Omega_2$ after propagation over a distance $z = L$ . Specifically, for an incident vortex beam with a topological charge of $l_1 = 1$, interaction with the medium prepared in the spatially dependent coherent superposition $\left|\psi\right\rangle = c_1\left|\phi_{210}\right\rangle + c_2 e^{-i\varphi}\left|\phi_{211}\right\rangle$ generates a new vortex field $\Omega_2$ with a topological charge of $l_2 = 2$ . By contrast, when the medium is prepared in the coherent superposition $\left|\psi\right\rangle = c_1\left|\phi_{210}\right\rangle + c_2 e^{i\varphi}\left|\phi_{21-1}\right\rangle$, the generated field carries a zero topological charge $l_2 = 0$, corresponding to a vortex-free output field. In the present model, the optical fields are assumed to copropagate along a common axis, and their directional separation is not explicitly considered. Consequently, the two fields may spatially overlap at the medium exit, and their separation may require frequency selective filtering, where applicable, or spatial mode sorting techniques, such as holographic filtering [30], cascaded Dove prism interferometry [31], and metasurface based beam sorting [32].

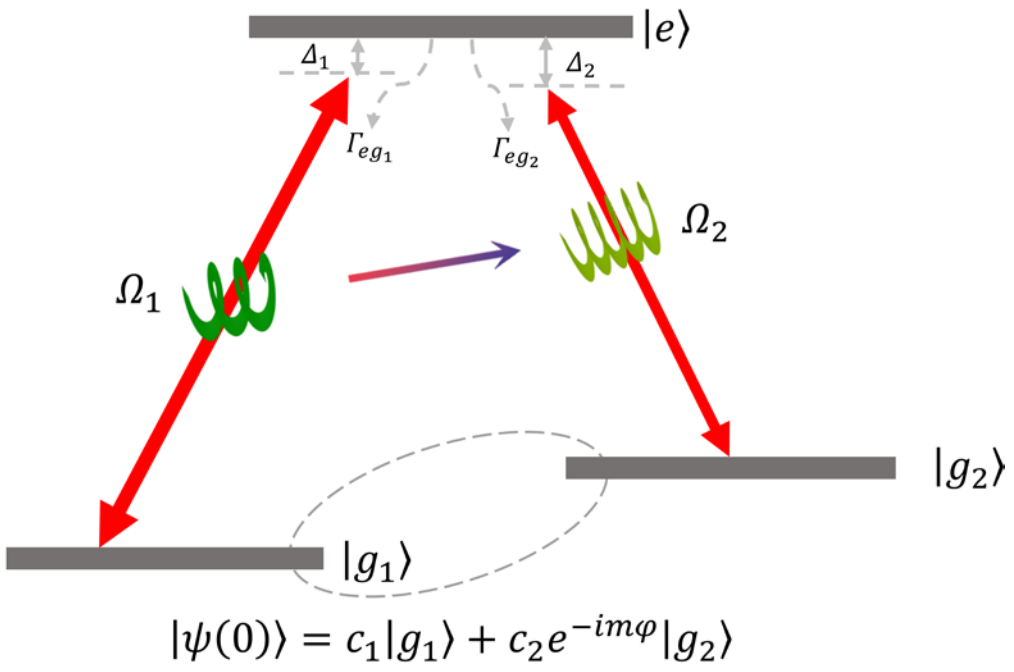

(a)

| | | |
|---|---|---|
| $\Omega_1$ Input | 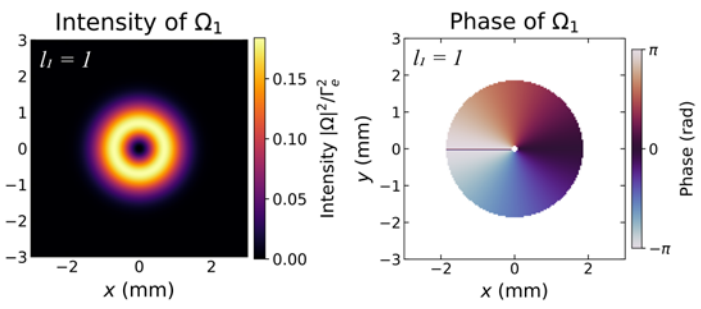 | |
| Superposition state | $\lvert\psi\rangle = c_1\lvert\phi_{210}\rangle + c_2 e^{-i\varphi}\lvert\phi_{211}\rangle$ | $\lvert\psi\rangle = c_1\lvert\phi_{210}\rangle + c_2 e^{i\varphi}\lvert\phi_{21-1}\rangle$ |
| $\Omega_2$ Output | 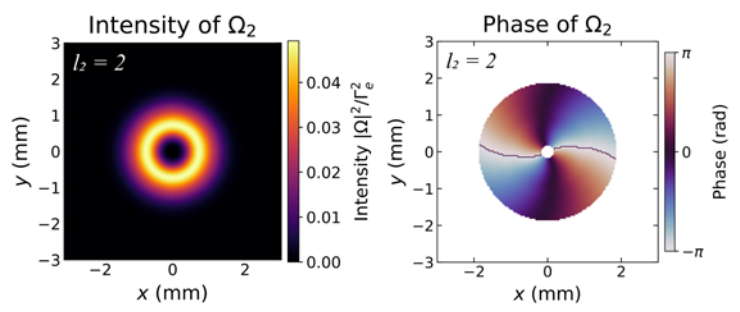 | 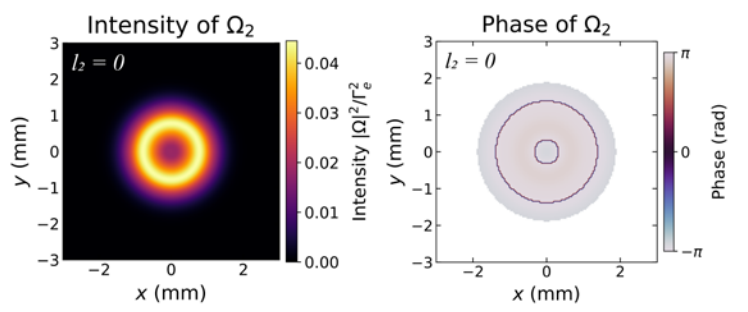 |

(b)

**Fig. 4.** Conversion between distinct OAM vortex modes. (a) Schematic illustration of the OAM mode conversion between vortex beams in a three-level Λ-type hydrogen atomic system. (b) Numerical results. The first row shows the intensity and phase distributions of the incident vortex beam, whose initial transverse profile is given by $\Omega_1(r,\varphi,0) = \Omega_0 (r/w_0)^{|l|} e^{-r^2/w_0^2} e^{il\varphi}$ with $w_0 = 1\text{mm}$ .The second row depicts the preparation of the atomic lower state into the corresponding coherent superposition state of the medium. The third row presents the intensity and phase distributions of the generated vortex field after propagation over a distance $z = L$ . All other parameters are the same as those used in Fig. 2.

## 5. Discussion

First, as demonstrated above, in a three-level Λ system, the direct generation of vortex light and conversion between vortex beams carrying different OAM modes can be achieved by preparing the two lower states in a spatially dependent coherent superposition. Following Ref. [23], the present treatment can also be generalized to analogous multilevel systems. In this case, the multiple lower states are prepared in a spatially dependent coherent superposition of the form

$$\left|\psi(0)\right\rangle = \sum_{j=1} c_j e^{-im_j\varphi}\left|g_j\right\rangle, \tag{24}$$

where $m_j$ is an winding number and $\varphi$ denotes the azimuthal angle about the $z$-axis. Such an extension could enable simultaneous control over multiple vortex light generation and conversion channels, thereby increasing the capacity for OAM mode processing and broadening the range of potential applications.

Second, we note a recent approach proposed by Li et al. [22], in which continuous-mode time-energy entangled W-class triphotons were directly generated through SSWM in a four-level triple-Λ atomic vapor system, with a substantially enhanced generation rate. By combining electromagnetically induced transparency with coherent control, their SSWM scheme enables multifunctional narrowband triphoton generation with long temporal coherence and controllable waveforms. The Supplementary Material of that work further discusses extensions to OAM and other degrees of freedom. Motivated by this work and recent studies [33], extending our scheme to spatially dependent coherent superpositions in multilevel systems may provide a route towards high-dimensional entanglement across multiple vortex light channels.

Finally, a possible experimental direction would be to explore spatially resolved rotational control in a coherently prepared hydrogen ensemble. Within the orbital approximation, magnetic field induced angular momentum dynamics could provide a possible starting point for implementing such local rotations [34]. The aim would be to imprint the required spatially varying helical relative phase onto a pre-existing coherent superposition. If such a coherence could be established and maintained, a weak input beam could probe the medium, while interferometric measurements could characterize the phase winding of the generated field. Similar considerations would apply to hydrogen-like ions. However, a spatially uniform rotation cannot generate the required transverse phase winding, and a concrete physical implementation capable of producing the full spatial coherence profile remains to be established. Accordingly, the present work should be regarded as a theoretical proposal. Detailed procedures for state preparation, spatial control, and coherence preservation remain to be developed and constitute important directions for future investigation.

## 6. Conclusion

In conclusion, we have proposed a theoretical scheme for vortex light generation and conversion in a coherently prepared three-level Λ medium. The key mechanism is the mapping of a prescribed azimuthally varying lower-state coherence onto the generated optical field. Within the model, this phase mapping leads to the topological charge relation $l_2 = l_1 + m$, where $l_1$ and $l_2$ denote the input and generated topological charges, respectively, and $m$ characterizes the coherence winding. Analytical results and numerical examples using hydrogen transition parameters illustrate vortex generation from a Gaussian input and conversion between distinct OAM modes. Under resonant conditions, dark-state interference strongly suppresses absorption during subsequent propagation after the initial bright-state component has been attenuated. This behavior arises because, for propagation distances $z \gg L_{abs}$, the system approaches a dark-state dominated regime associated with EIT. Nevertheless, absorption near the medium entrance and possible diffraction-related losses may still occur. Within the orbital approximation, the rotation relations for hydrogen atoms and hydrogen-like ions provide a formal description of relative-phase control. The proposed mechanism relies on the preparation and preservation of the prescribed spatial coherence, for which a complete experimental protocol remains to be established. Spatially resolved state preparation, coherence preservation, and extensions to multilevel systems for the simultaneous control of multiple OAM channels therefore represent important directions for future investigation. Together, these results provide a theoretical framework for manipulating optical OAM through spatially structured atomic coherence. This framework may offer a potential route for the generation and conversion of vortex modes and further advance our understanding of light-matter interactions involving optical OAM.

## Appendix

In this Appendix, we present the detailed derivation of the analytical solutions to Eqs. (8) and (9). We assume that only the field $\Omega_1$ is incident at the entrance of the medium, carrying a topological charge $l$:

$$\Omega_1(r,\varphi,0) = f_{in}(r)e^{il\varphi},\ \ \Omega_2(r,\varphi,0) = 0. \tag{A1}$$

For a Gaussian input beam,

$$f_{in}(r) = \Omega_0 e^{-r^2/w_0^2},\ \ l = 0. \tag{A2}$$

For a canonical vortex beam with topological charge $l$, the input field is chosen as

$$f_{in}(r) = \Omega_0\left(\frac{r}{w_0}\right)^{|l|} e^{-r^2/w_0^2}. \tag{A3}$$

Accordingly, the complete input field can be expressed as

$$\Omega_{LG}(r,\varphi,0) = \Omega_0\left(\frac{r}{w_0}\right)^{|l|} e^{-r^2/w_0^2}e^{il\varphi}. \tag{A4}$$

In cylindrical coordinates, the transverse Laplacian takes the form

$$\nabla_\perp^2 = \frac{\partial^2}{\partial r^2} + \frac{1}{r}\frac{\partial}{\partial r} + \frac{1}{r^2}\frac{\partial^2}{\partial\varphi^2}. \tag{A5}$$

For an arbitrary integer $n$, the following identity holds:

$$\nabla_\perp^2\left[u(r,z)e^{in\varphi}\right] = e^{in\varphi}\hat{B}_n^{(r)}u(r,z), \tag{A6}$$

where $\hat{B}_n^{(r)} = \frac{\partial^2}{\partial r^2} + \frac{1}{r}\frac{\partial}{\partial r} - \frac{n^2}{r^2}$ denotes the radial diffraction operator.

The optical fields can therefore be rigorously expanded as

$$\Omega_1(r,\varphi,z) = u_1(r,z)e^{il\varphi}, \tag{A7}$$

$$\Omega_2(r,\varphi,z) = u_2(r,z)e^{i(l+m)\varphi}. \tag{A8}$$

The expansions in Eqs. (A7) and (A8) follow directly from the rotational symmetry of the system. Consequently, Eqs. (8) and (9) can be rewritten as

$$\frac{\partial u_1}{\partial z} = \frac{i}{2k_1}\hat{B}_l^{(r)}u_1 - iX_1\left(|c_1|^2 u_1 + c_1^* c_2 u_2\right), \tag{A9}$$

$$\frac{\partial u_2}{\partial z} = \frac{i}{2k_2}\hat{B}_{l+m}^{(r)}u_2 - iX_2\left(c_1 c_2^* u_1 + |c_2|^2 u_2\right), \tag{A10}$$

where $\hat{B}_l^{(r)}$ and $\hat{B}_{l+m}^{(r)}$ denote the diffraction operators associated with the fields $\Omega_1$ and $\Omega_2$, respectively.

The corresponding initial conditions become

$$u_1(r,0)=f_{in}(r),\ \ u_2(r,0)=0. \tag{A11}$$

Applying the initial conditions (A11) to Eqs. (A9) and (A10) yields the exact analytical solutions for the radial amplitudes:

$$\begin{pmatrix} u_1(r,z) \\ u_2(r,z) \end{pmatrix} = \exp\left[ z\hat{A}_{l,m}^{(r)} \right] \begin{pmatrix} f_{in}(r) \\ 0 \end{pmatrix}, \tag{A12}$$

where the radial propagation operator is defined as $\hat{A}_{l,m}^{(r)} = \begin{pmatrix} \frac{i}{2k_1}\hat{B}_l^{(r)} - iX_1|c_1|^2 & -iX_1c_1^*c_2 \\ -iX_2c_1c_2^* & \frac{i}{2k_2}\hat{B}_{l+m}^{(r)} - iX_2|c_2|^2 \end{pmatrix}$.

Accordingly, the analytical solutions describing the propagation of the two optical fields in the medium are obtained as Eqs. (10) and (11).

## Acknowledgements

This work was supported by the National Natural Science Foundation of China (CN) Grant Nos. 12474353.

## Data Availability

The data that support the findings of this article are available from the authors upon reasonable request.

## Disclosures

The authors declare no conflicts of interest.